\documentclass[copyright,creativecommons]{eptcs}

\providecommand{\event}{FAVO 2009} 
\providecommand{\volume}{??}
\providecommand{\anno}{2009}
\providecommand{\firstpage}{1}
\providecommand{\eid}{??}
\usepackage{breakurl}    

\usepackage{rotating}
\usepackage{amsmath}
\usepackage{graphicx}

\newcommand{\usergoal}{detectPossibleOilspill(\lati,\longi,\area,\errorAccept)}
\newcommand{\goalone}{toBuy(satImage([\lati,\longi,\_,\area,\_,\sensorType,\_],\_))}
\newcommand{\goaloneN}{satImage([\lati,\longi,\_,\area,\_,\sensorType,\_],\_)}
\newcommand{\goaltwo}{toBuy(oilSpillDetect([\_,\_,\errorAccept],\_))}
\newcommand{\goaltwoN}{oilSpillDetect([\_,\_,\errorAccept],\_)}
\newcommand{\lati}{38.0}
\newcommand{\longi}{-9.4}

\newcommand{\freq}{5}

\newcommand{\area}{500}

\newcommand{\sensorType}{radar}

\newcommand{\errorAccept}{5}

\newcommand{\procOS}{$\langle procOSag,\{\langle provider(\goaltwoN), PC_4 \rangle\},\emptyset \rangle$}
\newcommand{\roleone}{$\langle requester(\goalone), PC_1 \rangle$}
\newcommand{\roletwo}{$\langle requester(\goaltwo), PC_2 \rangle$}
\newcommand{\rolethree}{$\langle provider(\goalone), PC_3 \rangle$}
\newcommand{\rolefour}{$\langle provider(\goaltwo), PC_4 \rangle$}

\def\init{\ensuremath{ag_{0}}}
\def\GI{\emph{identify goals}}
\def\NC{\emph{agree contract}}
\def\NW{\emph{agree Wf}}
\def\PD{\emph{discover partners}}
\def\PS{\emph{select partners}}
\def\ER{\emph{establish roles}}

\def\aids{\ensuremath{{AI}_{as}}}
\def\rids{\ensuremath{{RI}_{as}}}
\def\cids{\ensuremath{{CI}_{as}}}
\def\acl{\ensuremath{{ACL}_{as}}}
\def\onto{\ensuremath{{O}_{as}}}
\def\inte{\ensuremath{{L}_{as}}}
\def\soc{\ensuremath{\langle Agents, Services, Roles, Workflows, Contracts\rangle}}

\newtheorem{req}{Requirement}

\newcommand{\ntuple}[5]{<{#1},{#2},{#3},{#4},{#5}>}
\newcommand{\op}[1]{\; \xrightarrow{#1} \;}

\newcommand{\satImagein}{$[\lati,\longi,1000, 500, 5, optical,3]$}
\newcommand{\satImageout}{$results.data$}

\newcommand{\GItuple}{\ntuple{A_{init}}{G_{init}}{\emptyset}{\emptyset}{\emptyset}}
\newcommand{\PDtuple}{\ntuple{A_{pot}}{G_{init}}{\emptyset}{\emptyset}{\emptyset}}
\newcommand{\PStuple}{\ntuple{A_{pre}}{G_{init}}{\emptyset}{\emptyset}{\emptyset}}
\newcommand{\ERtuple}{\ntuple{A_{roles}}{G_{init}}{R_{vo}}{\emptyset}{\emptyset}}
\newcommand{\Nonetuple}{\ntuple{A_{vo}}{G_{vo}}{R_{vo}}{Wf_{vo}}{\emptyset}}

\title{A Formal Framework of 
Virtual Organisations as Agent Societies}
\author{Jarred McGinnis, Kostas Stathis
\institute{Department of Computer Science,\\
Royal Holloway, University of London, UK}
\email{
kostas.stathis@rhul.ac.uk}
\and
Francesca Toni
\institute{Department of Computing,\\
Imperial College London, UK}
\email{f.toni@imperial.ac.uk}
}

\def\titlerunning{VOs as Agent Societies}
\def\authorrunning{J. McGinnis, K. Stathis, F. Toni}

\begin{document}
\maketitle

\begin{abstract}
We propose a formal framework that supports a model of agent-based Virtual Organisations (VOs) for service grids and provides an associated operational model for the creation of VOs. The framework is intended to be used for describing different service grid applications based on multiple agents and, as a result, it abstracts away from any realisation choices of the service grid application, the agents involved to support the applications and their interactions. Within the proposed framework VOs are seen as emerging from societies of agents, where agents are  abstractly characterised by goals and roles they can play within VOs. In turn, VOs are abstractly characterised by the agents participating in them with specific roles, as well as the workflow of services and corresponding contracts suitable for achieving the goals of the participating agents. We illustrate the proposed framework with an earth observation scenario.
\end{abstract}

\section{Introduction}
\label{sec:intro}

The basic definition of Virtual Organisation (VO) is simple enough: organisations and individuals who bind themselves dynamically to one another
in order to share resources within temporary alliances~\cite{FosterKessel01}.
Several issues arise at various levels of abstraction when attempting to describe the alliance, the binding between members and the sharing of resources. 
We focus on an abstract, high-level description of VOs and their lifecycle
and define a formal model for VOs and their formation that can guide their realisation.
Like others (e.g. \cite{Matos,brainbrawn,patel-2005}) 
we focus on VOs that can be formed and managed automatically by 
intelligent agents. To support the envisaged automation agents {\em represent} organisations and individuals by providing and requesting resources/services, 
by {\em wrapping} them or by connecting to their 
published interface.
Agents are designed to incorporate the requirements of these organisations and individuals
and exhibit some {\em human aspects} while supporting the decision-making processes automatically.
Unlike existing work, we abstract away from concrete realisation choices of VOs
so that our models can be applicable to a range of service grid applications. Instead, the framework focuses on what we believe are likely to be the {\em essential elements} of VOs and ignore a number of lower-level aspects that are normally included in reference models for collaborative networks (see ~\cite{ecolead}).

Assuming a set of essential elements that are applicable across applications, our representation for the operational aspects of VOs  relies upon the notion of {\em VO life-cycle}, which reflects the orthodox  managerial and technical views of VOs, as proposed by~\cite{appel,gridbook2,websiteVO}. This lifecycle  
can be structured in 
three main phases: 
and the \emph{selection of partners}, 
formation,  operation, and dissolution.
In this paper we focus on the formation phase
with subphases:
(i)
\emph{initiation}, whereby an initiating agent \emph{identifies the goals} that it cannot achieve in isolation 
and \emph{discovers the potential partners} who can assist it in achieving those goals;  and
(ii)
\emph{configuration}, involving some form of \emph{negotiation}, trivial or otherwise, and the \emph{selection of partners}, 
trimmed down from those discovered, who will constitute the members of the VO once it is started.
We see a VO abstractly as a tuple consisting of \emph{agents} participating in the VO, \emph{roles} they play therein,
\emph{goals} the VO is set to achieve, 
the
\emph{workflow} of services being provided within the VO, and \emph{contracts} associated with these services, to which agents
are meant to conform.
We then  define the formation phase of a VO as a transition system between  
tuples providing partial approximations of the resulting VO.
Within our framework and for the purposes of VOs, agents are seen as existing within \emph{agent societies}:
we define these abstractly as our starting point. VOs then emerge within societies as a result of interactions amongst agents, as determined by the roles they play. 

Throughout the paper we shall exemplify
the proposed framework for VOs
by showing how it can be applied to the following simple but realistic \emph{earth obsevation scenario}. 

\begin{quote}
A government official is asked to investigate the detection of an offshore oil spill. As the ministry where the official works does not have direct access to earth observation facilities, the official typically follows a procedure. The first step of such a procedure is for the official to call a number of companies that control satellites which may provide suitable images. Satellites may have different orbits, sensors, capabilities and costs, so the official needs to discuss with satellite companies in order to select their most appropriate services for the task at hand. Satellites have names such as
\emph{Envisat}, 
\emph{ERS-1}, 
and
\emph{RADARSAT}.

As the satellite output is normally provided in the form of raw images, not immediately suitable
for the detection of an oil spill, the second step of the procedure involves the official calling companies that provide processing services by appropriate software, for example for 
\emph{format conversion} 
(into formats such as
TIFF, JPEG, etc),
\emph{reprojection} (into different coordinate systems),
\emph{pattern recognition} (to detect in the environment objects such as ships and buildings or features such as oil spills). After a post-processing image company is selected, the output of the satellite is processed by them and the resulting image will allow the official to identify the cause of the oil spill. 
\end{quote}
We will reinterpret this scenario by assuming that the government official is a user of an agent-oriented service grid. In such a grid, services such as oil spill detection and image processing are automatically discovered and negotiated by software agents that act on behalf of people and/or organisations. 
In this interpretation, the scenario will result into the formation of a VO that consists of the following parties:
the \emph{ministry official} and his agent acting as {\em service requester};
the  \emph{satellite company},
the \emph{post-processing company}, 
and their agents acting as {\em service providers}. The agents negotiate over the two requested services and orchestrate them into a {\em workflow} 
where the post-processing of the image data requires that the image data is created first.
To guarantee the properties and delivery of the services provided by the satellite company and the post-processing company
and to ensure those companies are compensated for their efforts, all the parties are involved into signing a
{\em contract}, which binds parties, in particular establishing their roles within the VO and defining a {\em Service Level Agreement} (SLA). 
An SLA specifies details of the service provision such as the
resolution of images, quality threshold, and time of delivery. Once the services are delivered via the execution of the workflow, the high-level goal of the user is satisfied and the VO is dissolved. 

The paper is organised as follows.
Section~\ref{sec:ag soc} presents the formalisation of the required abstractions, namely: agents, their roles and social norms specified as interaction protocols in an agent society, the services/resources available in that society, how these services can be combined in workflows, and how interactions in these workflows can be regulated by agreed contracts. These components will become the constituents of the VOs  and will be produced by a VO's formation phase. This phase is defined as a state transition system that will be formalised in section~\ref{sec:def}.
Finally, in section~\ref{sec:conc} we summarise our work and we outline our plans for the future.

\section{Agent Societies}
\label{sec:ag soc}

For the purposes of VOs, agents ``representing'' services
can be seen as existing within a {\em society} (of agents).
VOs emerge as a result of interactions amongst the agents in this society.
In other words, the agent society can be seen as the breeding environment~\cite{breeding}
for VOs. 
We will assume that an agent society exists prior to decisions and interactions leading to VOs.
However, 
typically this society is intended to be
``virtual'', in that
it is the implicit result of the existence of agents and services
within an agent-enabled grid/service-oriented architecture.

An agent society is characterised by roles that agents can adopt,
services available to and controlled by agents
in the society, possible combinations of these services within workflows,
and possible contracts between agents. 
Formally, 
$AgentSociety = \soc$. The elements of the $AgentSociety$ can be described as follows.

\begin{itemize}
\item $Agents$ is a (finite) set of agents, $\{A_1, \ldots, A_n \}$,
with $n \geq 2$; each agent is equipped with a set of individual goals,
an evaluation mechanism,
and a set of roles it can cover
(see section~\ref{sec:agents}). 
\item $Services$ is a (non-empty and finite) set of services represented by agents 
(see section~\ref{sec:services}). 
\item $Roles$ is a (non-empty and finite) set of roles that agents can play within the society as well as the
VOs, once they have been created. We require that
there are roles for $requester(s)$ and $provider(s)$  in $Roles$, for all $s \in Services$; roles are associated with interaction protocols (see section~\ref{sec:protocols}). 
\item $Workflows$ is a (non-empty) set of possible 
combinations of services in $Services$ (see section~\ref{sec:wf}).
\item $Contracts$ is a (non-empty) set of possible combinations of agents (in $Agents$),
roles (in $Roles$), 
and workflows (in $Workflows$) as terms in a contract (see section~\ref{sec:con}).
\end{itemize}
Note
that, in addition to roles for
$requester$ and $provider$ of all available services in the society,
$Roles$ may also include roles for
a $broker$ that provides information on how to obtain or use some services, an $arbitrator$ for making sure that
interactions for services are suitable regulated, and so on.
Finally, note that there are no goals of the agent society itself, and 
goals exist within agents only.
However,
VOs are goal-oriented:  we will see, in section~\ref{sec:def},
that the goals of VOs originate from those of individual agents.

The components of an agent society will be defined using
several abstract underlying languages. 
Here we single out these languages.
We adopt the following conventions:
variables start with capital letters; constants start with lower-case letters or numbers;
`\_' stands for
the anonymous variable. 

We 
use 
a set \aids{} of
{\rm agent identifiers}, that serve as unique ``names''
to address agents in the society, 
e.g. to support communication. 
An example is $satERS1Ag$, representing 
the satellite {\em ERS-1}.

We use a set 
\rids{}
of {\em role labels} for the definition of 
$Roles$.
We require that $requester(s)$ and $provider(s)$ belong to \rids, for all $s \in Services$.

We use {\em contracts identifiers},
\cids,  
univocally identifying and distinguishing contracts in $Contracts$.

We will assume some given, shared
{\em ontology} \onto,
which for simplicity we think of as a set of atomic and ground propositions.\footnote{
In general, \onto{} may need to contain hierarchical concepts,
for example a ``generic service'' may be defined as either
a ``satellite service'' or a ``processing service''.}
\onto{} will include 
(i) (an abstract representation of) all services in $Services$, 
(ii) generic infrastructure knowledge,
e.g. for querying registries holding information about agents and the services
they provide.
An example of the latter may be
$provides(X,satImage(in,out))$ instantiating $X$ to $satERS1Ag$, representing that the agent named $satERS1Ag$ represents
a  provider of service $satImage(in,out) \in Services$.

We will see that VOs emerge in an agent society by communicative interaction amongst its members. As usual, communicating agents will share a {\em communication language} that will act as a ``lingua franca'' amongst agents.
We thus assume as given a language \acl{} 
of locutions.
As conventional, 
locutions consist of
a {\em performative} and a {\em content}.
Examples of locutions in \acl{} may be
$request(s)$ and $accept(s)$,  where $s \in Services$ is the content. 

Each individual agent is equipped with an {\em internal language}
to express its knowledge/beliefs and goals. Since
the goals of VOs
are derived from the individual agents' goals,
we need to assume that the agents share at least a fragment
of their internal languages.
This fragment can be also used to express, e.g.,
conditions in protocol clauses (see section~\ref{sec:protocols}).
We will refer to this shared fragment of all agents' internal languages as
\inte.
We require that the sentence
$true$ is contained in this language,
as well as sentences built using 
the usual connectives $\wedge$ and $\neg$. We assume that this language is propositional.
Examples of sentences in
\inte{} may be
$toBuy(satImage(in,out))$.
Sentences are meant to be evaluated using the agents' internal evaluation mechanisms
(see section~\ref{sec:agents}).

Note that there are no eligibility conditions to choose which agents enter the society, as we assume an open setting where agents can freely circulate. In this context, VOs provide a mechanism for defining which agents can be suitably put together to help solving each others' goals.

\subsection{Services}
\label{sec:services}

{\em Concrete services}, that can be executed by their providers,
are described using sentences in \onto.
Examples of concrete services 
are $satImage(in,out)$ with $in$ and $out$ representing the inputs and outputs for the service (e.g.  $in$ may be
\satImagein{} and $out$ may be \satImageout)
\footnote{Here, 38.0 and -9.4 are the latitude and longitute coordinates of the area
to be scanned, 1000 is the resolution of the image in metres,
500 is the km$^2$ area to be scanned,
5 is the frequency in hours for the area to be scanned,
$optical$ is the type of sensor to be used, 3 is the wave frequency to be used in the scan,
\satImageout{} is the name of the file produced by
Envisat.}
and 
some service for detecting oil-spills $detectOilSpills([a,5],b)$\footnote{
Here,  $a$ represents
the input raw satellite data,
5 is the acceptable threshold for false positives
and $b$ is the output processed data image,
as computed by the provider of $detectOilSpill$.
Note that 
algorithms for detecting oil spills may
occasionally give false positives, namely  indicate that there is an oil spill 
at some location where
in reality no oil spill is present there.
The lower the acceptable false positive threshold requested from a service,
the more expensive the service.}.

In order to accommodate negotiation for the provision of (concrete) services during the formation of VOs,
it is useful that agents are able to talk about {\em partially uninstantiated} and \emph{abstract} services,
before they commit to any concrete instantiation
(actually, it may happen that this instantiation  can only be provided at
the time of execution of the services).
For example,
an agent may require, for some given $a$,
$detectOilSpill([a,T],B)$,
where the threshold $T$ and the output processed data image $B$ are as
yet unspecified ($T$ may be associated with constraints, e.g. $T \geq 5$).

In summary, we adopt the following description of services:

\begin{center}
\begin{tabular}{lll}
$serviceName(In,Out)$ & : & uninstantiated service ({\em abstract service});\\
$serviceName(in,Out)$ & : & partially instantiated service;\\
$serviceName(In,out)$ & : & partially instantiated service;\\
$serviceName(in,out)$ & : & fully instantiated service ({\em concrete service});\\
$serviceName$         & = & predicate
, with $serviceName(i,o) \in \onto$, \\
&& for
constants or lists of constants $i,o$; \\
$in,out$               & = & constants or lists of constants; \\
$In,Out$               & = & variables or lists of variables. \\
\end{tabular}
\end{center}

The $serviceName$ can be seen as the ``type'' of service being provided by $serviceName(in,out)$.
We will often refer to an abstract service
$serviceName(In,Out)$ simply as $serviceName$.
Also, an abstract or partially instantiated service can be seen logically
as representing a disjunction of concrete services (one for each possible instantiation).
We could thus see the process of negotiating an instantiation of
an abstract or partially instantiated service as
the process of negotiating a concrete service in a set of alternatives (the disjuncts).

For our scenario, we need four types of services, namely
$satImage$ 
and three 
processing services (with $serviceName$ one of $formatConversion$, $reprojection$ and $detectOilSpill$).
We have already seen examples of concrete services of type
$satImage$ and $detectOilSpill$. 

\subsection{Roles and Protocols}
\label{sec:protocols}

A role is defined as a tuple $\langle rid , PC \rangle$ where
$rid \in \rids$ is the label of the role, and
$PC$ is a {\em Protocol Clause},
understood in this paper as 
a (non-empty and finite) set of {\em Operations}
defined as follows: 

\begin{center}
\begin{tabular}{llll}
$Operation$ & = & $\psi [send(m,i,rid')] \phi$& (send operation)\\
     & $|$ & $\psi [receive(m,i,rid')] \phi$& (receive operation)\\
$\psi$ & $\in$ & $\inte \cup \onto$ & (precondition)\\
$\phi$ & $\in$ & $\inte $ & (postcondition) \\
$m$ & $\in$ & $\acl$  & (locution) \\
$i$    & $\in$ &  \aids & (unique identifier of agent)\\
$rid' $ & $\in$ &  \rids & (role label) 
\end{tabular}
\end{center}

Intuitively, 
each operation 
has three parts:
a locution $m$ in $\acl$,
an identifier $i$ in \aids{} of the communicative partner
(i.e. the intended recipient  or the 
actual sender of message $m$, respectively for $send$ and $receive$),
and the identifier $rid'$ of the role that the agent $i$ is intended to be playing
when receiving or sending the message (respectively for $send$ and $receive$).
An agent
can send or receive the locution (according to what the operation specifies)
if and only if the evaluation mechanism of the agent 
evaluates the precondition $\psi$
to true.
Once the message is sent or received, the postcondition $\phi$
\emph{will automatically hold} (namely the evaluation mechanism of the agent
will evaluate this condition positively after the message is sent or received,
until further changes).
Moreover, when an agent
$i'$
playing some role $\langle rid,PC\rangle$
sends a locution 
$send(m,i,rid')$
to some other agent
$i$, 
$i$ receives the message from $i'$ indicating that $i'$ sent it while playing role $rid$:
$receive(m,i',rid)$.
This message will be ``processed'' by $i$
using the role with identifier $rid'$.

We could adopt other formalisms for communication, e.g. non-determinisitc finite-state automata.
The reason we have chosen protocol clauses is that this formalism is an abstraction of several existing formalisms
for modelling inter-agent communication,
e.g. LCC \cite{dave:iclp} and dialogue constraints \cite{atal01:negotiation}.

To illustrate roles,
consider a simple example where an agent playing the role of $requester$
(of some service $S$)
sends a $request$ to an agent it believes
to be a provider of $S$, and requiring it to be playing the role of $provider$ of $S$.
The $provider$ agent replies with $accept$  or $refuse$
depending on whether it is indeed a provider of that service $S$ (and it wants to sell that service)
or not (respectively).
\vspace*{0.2cm} 

{\em
\begin{tabbing}
\=$\langle$ requester(S), \= \{ \= \\
\>\>\>toBuy(S) $\wedge$ provides(Ag,S) $[$send(request(S), Ag, provider(S))$]$ requested(S,Ag),\\
\>\>\>requested(S,Ag) $[$receive(accept, Ag, provider(S))$]$ bought(S),\\
\>\>\>requested(S,Ag) $[$receive(refuse, Ag, provider(S))$]$ true\\
\>\> \} \\
\>$\rangle$
\end{tabbing}}

{\em
\begin{tabbing}
\= $\langle$ provider(S),\= \{ \= \\
\>\>\>true $[$ receive(request(S), Ag, requester(S)) $]$ requestedBy(Ag,S),\\
\>\>\>requestedBy(Ag,S)$\wedge$ toSell(S) $[$ send(accept, Ag, requester(S))$]$ sold(S), \\
\>\>\>requestedBy(Ag,S)$\wedge$ $\neg$ toSell(S) $[$ send(refuse, Ag, requester(S))$]$ true\\
\>\>\}\\
\>$\rangle$
\end{tabbing}}

\noindent In the two {\em protocol clause schemata} above variables {\em S} and {\em Ag} are used instead of concrete values.
These variables are implicitly universally quantified over the appropriate languages.

A protocol clause for a role
defines the communicative actions for any agent
adopting the role. 
However, protocol clauses typically relate to other protocol clauses
and give a global picture of the interaction amongst agents and roles.
For the earlier example,
the two roles,
$requester(S)$ and $provider(S)$,
are related to one another to form a simple \emph{negotiation protocol}.
Intuitively,
a {\em protocol} is a (non-empty and finite) set of protocol clauses
for roles in $Roles$ that are ``related'' to one another,
possibly indirectly.

With an abuse of notation, we will often refer to
a role simply by its identifier
and will use the identifier to stand for the corresponding protocol clauses.
Moreover, 
when an agent needs to play the role of
$requester$ for any service,
we use 
$requester$ to stand for $requester(s)$ for any service $s \in Services$.
Finally, we use $provider(serviceName)$ 
to indicate that a service provider 
can provide all instances of an abstract service
$serviceName(In,Out)$ or
when we are interested in the
provision of some instance of this service without
specifying which one.

\subsection{Agents}
\label{sec:agents}

For the purposes of VOs, an agent can be seen abstractly as a tuple
$\langle i,R,G \rangle$
where
$i \in \aids$ is the unique identifier for the agent;
$R\subseteq Roles$ is a (non-empty) set of roles that the agent can play within the society;
$G\subseteq \inte$ is a (non-empty) set of goals of the agent.

An agent is also equipped with
an {\em evaluation mechanism} for determining
whether
(i) preconditions in roles are satisfied,
(ii) goals are fulfilled by the agent in isolation.
This mechanism is affected by the execution of protocols in that
postconditions of protocol clauses are taken into account by this evaluation mechanism
(they are satisfied after the protocol clause is executed, until overwritten by further postconditions).
We do not include this evaluation mechanism within the representation of an agent
in a society as this mechanism is private to agents
and different
agents may adopt different such mechanisms in general.
 
Roles and goals of an agent
$\langle i,R,G \rangle$
are inter-related as follows:
\begin{itemize}
\item[(a)]
$\forall r \in R$, $\exists g \in G$ which ``enables'' $i$ to adopt $r$,
namely 
the need to fulfil $g$ is a precondition for
one of the protocol clauses in $r$;
\item[(b)]
$\forall g \in G$, $\exists r \in R$ such that playing the role $r$
gives $i$ a ``chance of fulfilling'' $g$
namely 
one of the protocol clauses in $r$ admits the fulfilment of $g$ as one of its postconditions.
\end{itemize}

\noindent As an example, consider the earlier protocol clause for the role $requester(S)$ where
{\em toBuy(S)} corresponds to a goal
and {\em bought(S)} corresponds to the goal being fulfilled.
Example agents for our scenario are  
\vspace*{0.2cm}

$\langle clientAg,\{requester\},\{toBuy(someI), toBuy(someD)\} \rangle$ 
 
$\langle satERS1Ag,\{requester(detectOilSpill),
provider(satImage)\}$, $\{toSell(someI)\} \rangle$ 

$\langle detectAg,\{provider(detectOilSpill)\},\{toSell(someD)\} \rangle$ 

\vspace*{0.2cm}
\noindent where $someI$ is of the form $satImage(in,Out)$ 
and $someD$ is of the form $detectOilSpill([Out,t],Out')$
for some $in$ and $t$ (as discussed earlier).
Here, 
the agent identified by $clientAg$ represents the government ministry
and can only play the $requester$ role (for any service),
the agent  identified by $satERS1Ag$ represents the {\em ERS-1} satellite
and can play both the $requester$ role for $detectOilSpill$ services
and the $provider$ role for $satImage$ services,
and the    agent  identified by $detectAg$ can play only the $provider$ role
for $detectOilSpill$ services.
The agents' goals allow them to engage in interactions 
following the protocols for the roles they are equipped with 
(see the simple protocol clauses of section~\ref{sec:protocols}).

\subsection{Workflows}
\label{sec:wf}

We see workflows simply as (non-empty) sets of services\footnote{More generally,
workflows can be compositions, e.g. by sequencing or parallel execution, of services.} 
possibly annotated with ``constraints'',
that are sentences in \inte.
Services may be abstract, partially instantiated or concrete, as in section~\ref{sec:services}.
As an example, consider 
the annotated workflow (consisting of a single partially instantiated 
service)

\vspace*{0.2cm}
$\{satImage([\lati,\longi,Res,\area,\freq,ST],Out)\}$ $\bigcup$ 
$\{Res \in [900,1100], ST \in \{radar, optical\}\}$

\vspace*{0.2cm}
\noindent where the resolution $Res$ and sensor type $ST$ arguments
are constrained within the annotation. 

We require that the constraints annotating workflows are satisfiable in \inte.
Annotations only make sense for workflows with at least one partially instantiated or abstract service.
They are intended to restrict the possible instantiations of the services in
the workflow. Typically, as in the example above,
they pose restrictions on the variables occurring in the
services of the workflow.

We will adopt the following terminology:
an {\em abstract workflow}
is a workflow with at least one abstract or partially instantiated
service (with or without annotations);
a {\em concrete workflow} is a workflow consisting solely of concrete services (without annotations).
Also, a concrete workflow may or may not be executable,
and that, prior to execution of a workflow, may need to be appropriately
set up. 
In this paper, we focus on the {\em formation} phase of VOs and ignore
execution issues that may occur in the {\em operation} phase.

\subsection{Contracts}
\label{sec:con}

Inspired by
web service contract standards~\cite{wsagree},
we define a contract as
$\langle Cid,Context,SDT,GT \rangle$
where

\begin{itemize}
\item
$Cid \in \cids$ is a unique identifier for the contract;
\item
$Context$ indicates all agents bound by the contract (the ``contracted parties'')
and their role in the contract,
formally seen as a set of pairs of the form $(AgentId,AgentRole)$
such that
$\langle AgentId, R, \_ \rangle \in Agents$ and
$AgentRole \subseteq R$;
\item
$SDT$, the {\em Service Description Terms},
is a 
workflow, consisting of 
services being contracted; 

\item
$GT$, the {\em Guarantee Terms}, is a (possibly empty)
set of sentences in \inte{} that define
the assurances with regards to the services described in $SDT$.

\end{itemize}
The $GT$ component of a contract
may also include rewards/penalties for the contracted parties
and refer to roles (and protocols) to be used by agents in the case of exceptions. For example, if blame for failure is disputed, there may be a clause in $GT$
defining a protocol for arbitration.

By definition of $Context$, we require that the contracted parties play, within the contract,
some of the roles they are equipped with.
We require that there are at least two different agents involved in any contract,
and that there is at least one agent
playing the role of $requester(s)$ and at least one agent
playing the role of $provider(s)$ for some service $s$,
namely: 
\begin{itemize}
\item 
$\exists (id1,role1), (id2,role2) \in Context$ such that
$id1 \neq id2$, and
$requester(s) \in role1$,
$provider(s) \in role2$.
\end{itemize}

\noindent We exclude the possibility that the same agent may be a provider and a requester for the
same service: 

\begin{itemize}
\item
$\not \exists (id,role) \in Context$ such that,
for some $s\in Services$,
$\{requester(s), provider(s) \} \subseteq role$.
\end{itemize}

\noindent We require that for all the services in $SDT$
there exists an agent in $Context$ providing that service: 

\begin{itemize}
\item 
$\forall s\in SDT$,
$\exists (id,role) \in Context$ such that
$role=provider(s)$.
\end{itemize}

\noindent A simple example of a contract is:
{\em
\begin{tabbing}
\= $\langle$ co\=ntr\=actX,\\
\>\> \{$\langle$clientAg,\{requester(formatConversion)\}$\rangle$, 
$\langle$procF, \{ provider(formatConversion)\}$\rangle$\},  \\
\>\>\{formatConversion([image.jpeg, jpegTOgif], imageGIF.gif)\},\\
\>\>\{dueBy(ImageGIF.gif,1400hrs,12.4.09), priceReduced(ImageGIF,1400hrs,12.4.09,reduction(0.5))\}\\ 
$\rangle$ 
\end{tabbing}}

\begin{sloppypar}
\noindent The above contract, identified as $contractX$,
is between $clientAg$ and $procF$ for the delivery  of 
(an instance of) the service
$formatConversion$, for converting the file $image.jpeg$ 
using the operation called $jpegTOgif$.
The service has a due delivery date specified using 
the 
$dueBy$ 
predicate.
The clause on $priceReduced$ 
specifies that
the price charged will be halved  
if the provider 
fails to deliver
on time.
\end{sloppypar}

\subsection{From Agents and Services to the Agent Society}

Given 
$Agents$
as in section~\ref{sec:agents}
and 
$Services$ as in section~\ref{sec:services}
an agent society
``emerges'' with:
\begin{itemize}
\item $Roles = \bigcup_{\langle i,R,G \rangle  \in Agents} R$
(the possible roles
are all roles that agents within the society can play);

\item the concrete workflows in $Workflows$ 
are all possible (non-empty) sets of services,
namely $(2^{Services} - \emptyset) \subset Workflows$, while
the remaining elements of $Workflows$ are abstract, 
possibly annotated ``versions'' of these concrete workflows; 

\item $Contracts$ is built solely from elements of $Workflows$, $Roles$ and $Agents$.
\end{itemize}

\noindent We require also that each service is ``represented'' by an agent within the society,
in other words the possible services in the society
correspond to all services the agents can provide:
\begin{itemize}
\item 
$\forall s \in Services$,
$\exists \langle provider(s),\_ \rangle \in Roles$
\end{itemize}

\noindent However, it may be the case that several alternative protocols exist in the society for the same role, namely:
$\langle rid, PC\rangle$ and $\langle rid, PC'\rangle$ both belong to $Roles$
for $PC \neq PC'$.
The creation of a VO will need to address the choice of protocols being used for negotiation of workflows and contracts.

\section{The VO Formation Phase}
\label{sec:def}

VOs can be defined as tuples 
$\ntuple{A_{vo}}{G_{vo}}{R_{vo}}{Wf_{vo}}{Con_{vo}}$
whose components can be described as follows.

\begin{itemize}
  \item 
$A_{vo} \subseteq Agents^*$ with
$Agents^*$ = $\{\langle i, R', G'\rangle | \langle i, R, G \rangle \in Agents$ and
$R' \subseteq R$, $G' \subseteq G\}$.
$A_{vo}$ contains at least two agents and 
exactly one agent in  $A_{vo}$ is referred to as the {\em initiating agent}, which is
denoted \init.

  \item $G_{vo}$ is a set of goals for the agents in $A_{vo}$,
which contains at least a goal of the initiating agent:
$G_{vo} \subseteq \bigcup_{\langle i,R,G \rangle \in Agents} G$
and $G_0\cap G_{vo} \neq \emptyset$, where 
$\langle \init,R_0,G_0 \rangle \in Agents$.

  \item $R_{vo}$ is a set of roles to be played by the participating agents, 
where $R_{vo} \subseteq Roles$.

  \item $Wf_{vo} \subseteq Workflows$ is the workflow of services
currently agreed amongst the agents in $A_{vo}$.
  \item $Con_{vo} \subseteq Contracts$ is a set of contracts between the agents in
$A_{vo}$.
\end{itemize}
$Agents^*$ represents the set of all possible ``full'' or ``partial'' 
specifications of agents,
corresponding to concrete choices of roles agents may play and goals they may focus on within a specific VO.
$A_{vo}$ describes the (partial specifications of) agents involved in the VO,
as providers or requestors of services,
or in whichever other roles, as indicated by
$R_{vo}$. 
$A_{vo}$ only describes the agents insofar as their involvement in the VO
is concerned (and thus possibly omitting some of their goals and roles, 
not relevant to the VO).
The $Wf_{vo}$ and $Con_{vo}$
components 
determine the behaviour of the VO and its members during 
the execution and dissolution phases of VOs.
The $G_{vo}$ and $V_{vo}$ components are related to the 
the $A_{vo}$ component in that 
 $G_{vo} = \bigcup_{\langle i,R,G \rangle \in A_{vo}} G$
 and $R_{vo}= \bigcup_{\langle \init,R_0,G_0 \rangle \in A_{vo}} R$. 

In our model, a VO 
is instantiated during the formation phase,
through interactions amongst the agents composing the VO.
These interactions can be understood in terms of several 
operations  
progressively refining ``partial'' representations of VOs.
These operations are defined as transitions,
as outlined below.
In the remainder,
$Ids(A)$ refers to all identifiers of (partial specifications of) agents in the set $A$.

\subsection{Goal Identification} 

The \GI{} transition results in the additions of the initiating agent 
\init{} 
and (some of) its goals into the partial VO tuple.
These are goals that the agent cannot achieve on its own.
Given
\begin{center}
$\ntuple{\emptyset}{\emptyset}{\emptyset}{\emptyset}{\emptyset} \op{identify \; goal} \GItuple$,\\ 
\end{center}
\noindent then 

\begin{itemize}
\item $A_{init} = \{\langle \init, \emptyset, G_{init} \rangle \}$
for some $\langle \init, R_0 , G_0\rangle
\in Agents$
such that some goals $G_0' \subseteq G_0$ cannot be fulfilled 
by \init{} in isolation;

\item  
$G_{init} = G_0'$. 
\end{itemize}

\noindent Here, goal fulfilment is
determined using the evaluation mechanism of agent \init{} (see section~\ref{sec:agents}).
Note that no role is yet identified for \init: this will be done in
transition \ER{} (see section~\ref{sec:em}).

In order to ground this transition to our scenario, we assume that an agent  
$clientAg$ is informed by its user that a possible oil spill has been reported by a passing
vessel. The user provides the following information to the agent: latitude, longitude, acceptable false positive threshold and scan area. Given the user's parameters the agent initiates the VO formation process 
by first identifying its goals.
The parameters correspond to high-level goals, 
that will later be decomposed into specific workflows.
In the example, the high-level goal 
$\usergoal$ given by the user is to detect an oceanic 
oil spill off the Portuguese coast 
at a latitude and longitude of \lati{} and \longi{} 
with an acceptable false positive threshold of \errorAccept\%
for the surrounding \area{} square kilometres.
The agent may decide that to satisfy this high-level
goal it needs two services: 

\hspace*{0.1cm} 

$g_1=\goalone$, and 

$g_2=\goaltwo$ 

\hspace*{0.1cm} 

\noindent where the sensor-type for the satellite providing the image
must be 
radar, because of the required resolution and weather conditions,
and  
once this image data is obtained a service is needed 
to provide the actual identification of the oil spills on the images. 
In summary, \GI{} will compute
$A_{init} = \{\langle clientAg , \emptyset , G_{init} \rangle \}$ and 
$G_{init} =  \{g_1, g_2\} $. 

Note that in our model goals of VOs emerge from goals of agents. Once the goals of VOs have been identified, they will dictate the behaviour of agents during the operation of VOs.
 
\subsection{Partner Discovery}

The \PD{} transition results in the addition of a number of agents to the
set of agents in the current partial VO (after \GI).
Whether by traditional means such as a yellow page 
registry or through `friend of a friend' discovery utilising the multiagent system, 
the VO tuple is transformed to include potential partners that the initiating agent believes will help it reach its goals, notably by providing suitable services.
Given 
\begin{center}
$\GItuple \op{discover \; partners} \PDtuple $,
\end{center}
\noindent then
$A_{pot} = A_{init} \bigcup A_{queryresult}$, where 
$A_{queryresult}$ includes those potential partners such that 
\begin{itemize}
\item $A_{queryresult} \subseteq Agents^* - A_{init}$
and each element in $A_{queryresult}$ is of the form $\langle i, \emptyset, \emptyset \rangle$;
\item each agent in $A_{queryresult}$ is a provider of one of the services 
in $G_{init}$ 
namely,
for each
$i \in Ids(A_{queryresult})$,
if $\langle i, R, G \rangle \in Agents$
then
$\exists s$ such that $toBuy(s) \in G_{init}$ such that
$\langle provider(s), PC \rangle \in R$.
\end{itemize}

In our example, $clientAg$ finds that two satellite image providers 
advertise the services it is interested in. Both agents $satERS1ag$ and $radSatAg$ represent a 
radar-capable
polar satellite with orbits amenable to the area of interest. 
Moreover, there is one agent, $procOSAg$, 
who can provide the type of image processing in which $clientAg$ is interested.
After this transition is completed, 

$A_{queryresult} = \{ \langle satERS1ag,\emptyset,\emptyset \rangle, \langle radSatAg,\emptyset,\emptyset \rangle, \langle procOSAg,\emptyset,\emptyset \rangle \}$.

\subsection{Partner Selection}

The set of potential partners discovered by \init{} 
may include
unreliable or untrustworthy ones.
The \PS{} transition allows the agent to prune the results of the \PD{} transition.
We do not provide a detailed specification of the pruning mechanism needed to support this stage as this is largely dependent on mechanisms for assessing trustworthiness and reliability. Several such mechanisms exist in the literature. Any could be used here.

Generally, given
\begin{center}
$\PDtuple \op{select \; partners} \PStuple$, 
\end{center}

\noindent then 

\begin{itemize}
\item
$A_{pre} \subseteq A_{pot}$;
\item $\init \in Ids(A_{pre})$;
\item
for each $s$ such that 
$toBuy(s) \in G_{init}$ 
there exists

$ i \in Ids(A_{pre})$ such that

if $\langle i, R, G \rangle \in Agents$
then

$\exists \langle provider(s), PC \rangle \in R$.
\end{itemize}

In the example, after the \PS{} transition is completed:
\begin{quote}

$A_{pre} = \{\langle clientAg , \emptyset , G_{init} \rangle, \langle satERS1ag,\emptyset,\emptyset \rangle, \langle procOSag,\emptyset,\emptyset \rangle \}$. 
\end{quote}

Note that, in general, several providers of the same service may still be
in $A_{pre}$ after the pruning.

\subsection{Establish Roles} 
\label{sec:em}

Before the agents are able to negotiate workflows and contracts,
the roles they will be playing in the negotiation
need to be established.
These roles (with their protocols)
are the social norms used for forming the VO.
Establishing these roles also amounts to establishing protocols for them
(as our roles include protocols).
Roles include requester and provider roles, but may also include
other roles (e.g. that of arbitrator, or contract-negotiator if agents
other than provider and requester agents may be needed to support the contract negotiation stage of VO formation).
For simplicity, we assume that these roles are decided by the initiating agent,
and that, given
\begin{center}
$\PStuple \op{establish \; roles} \ERtuple$,
\end{center}
then

\begin{itemize}
\item $Ids(A_{pre}) \subseteq Ids(A_{roles})$;
\item $A_{roles} = A_{pre}^* \cup A_{rest}$, where
\begin{itemize}
\item $A_{pre}^* =
\bigcup_{\langle i, \emptyset, G \rangle \in A_{pre}}
\{\langle i, R_i, G_i \rangle\} $ for some $R_i$s such that
$R_i \subseteq R^*_i$ and $G_i \subseteq G_i^*$ where $\langle i, R^*_i, G_i^* \rangle \in Agents$;
\item $A_{rest} \subseteq Agents^*$  (where $A_{rest}$ may be empty);
\item $A_{rest} \cap A_{pre}^* = \emptyset$;
\end{itemize}

\item $R_{vo} = \bigcup_{\langle i, R_i, \_ \rangle \in A_{roles}} \{R_i\}$;
\item for each $s$ such that $toBuy(s) \in G_{init}$,
there exists exactly one $\langle provider(s), PC_{provider(s)} \rangle$
and exactly one  $\langle requester(s), PC_{requester(s)} \rangle$ in $R_{vo}$;
\item for every $\langle r_1, PC_{r_1} \rangle$
and $\langle r_2, PC_{r_2} \rangle$ in $R_{vo}$,
if $r_1=r_2$ then $PC_{r_1}=PC_{r_2}$,
namely there is exactly one role for each role identifier in $R_{vo}$.
\end{itemize}

Note that we do not impose that \init{} plays the role of requester of all
the services in the workflow it wants to instantiate:
indeed, in general it may be possible that
\init{} delegates the task of requesting some or all services to some other agent.
In particular, it may be the case that 
$\langle \init, \emptyset, G_{init} \rangle$ belongs to $A_{roles}$.
Also, we allow for the same agent to play several roles in a VO
(namely, $\langle i, R_i, G_{i} \rangle$ may belong to $A_{roles}$
with $R_i$ containing more than one role).

In our running example, once the \ER{} stage is completed: 

\begin{quote}
$A_{roles} = \{$
$\langle clientAg, \{\langle requester(\goaloneN), PC_1 \rangle,$\\
\hspace*{3.4cm}$\langle requester(\goaltwoN), PC_2 \rangle \}, G_{init} \rangle, $\\
\hspace*{1.6cm}$\langle satERS1ag, \{\langle provider(\goaloneN), PC_3 \rangle \},\emptyset \rangle$\\
\hspace*{1.6cm}\procOS $\}$

$R_{vo} = \{$ 
\roleone, \\
\hspace*{1.3cm}\roletwo, \\
\hspace*{1.3cm}\rolethree, \\
\hspace*{1.3cm}\rolefour
$\}$
\end{quote}
Here, the $PC_i$ are protocol clauses that the agents commit 
to follow during the negotiation of workflows. 
In this specific example, no role/protocol is specified for the \NC{} transition.
Note that other agents may be brought 
into $A_{roles}$ at this stage to play these new roles.

\subsection{Negotiation}
The negotiation activities in the VO formation amount to
1) agreeing a concrete workflow (\NW) and
2) agreeing a set of contracts amongst agents contributing to the workflow, by providing services in it,
and the initiating agents (stage \NC).
Both transitions make use of roles (and protocols) identified at the \ER{} transition:
communicating by following these protocols
agents agree on the provision of services and contracts.
Negotiation may result in additional goals to be added,
as goals of agents providing services.
The \NC{} transition may cause no changes in the partial VO tuple,
if no suitable roles have been
computed by the \ER{} transition.
For lack of space we will only describe the \NW{} transition.

In order for the computed VO to be meaningful, it needs to compute
a workflow that is concrete or
partially instantiated, but can be fully instantiated when the workflow is executed.
This workflow instantiates the abstract workflow corresponding to 
the $toBuy$ goals in $G_{init}$.
This instantiation may be obtained after several negotiations,
each following the protocols of the roles identified after the \ER{} transition,
each resulting in a service becoming instantiated.
After each instantiation,  the initiating agent puts those instantiated services into the workflow component of the VO tuple.

Generally, given\\
\begin{center}
$\ERtuple \op{agree \; Wf} \Nonetuple$, 
\end{center}

then 

\begin{itemize}
\item  $Ids(A_{vo}) \subseteq Ids(A_{roles})$;
\item  $\init \in  Ids(A_{vo})$;
\item  for each $s$ such that $toBuy(s) \in G_{init}$ 
there exists exactly one agent $i \in Ids(A_{vo})$
such that $\langle i, R_i, G_i \rangle \in A_{vo}$
and $provider(s) \in R_i$,
and a successful dialogue between \init{} and this agent $i$
with  \init{} playing the role of
$requester(s)$ and $i$ playing the role of $provider(s)$;

\item  for each $\langle i, R_i, G_i \rangle \in A_{vo}$,
if $\langle i, R_i^*, G_i^* \rangle \in A_{roles}$ then
$R_i^*=R_i$ (namely roles cannot be changed at this stage);

$G_i \supseteq G_i^*$ (namely goals can only be added at this stage);

if $\langle i, R_i^{**}, G_i^{**} \rangle \in Agents$ then
$G_i \subseteq G_i^{**}$ (namely all goals are chosen from the pool of goals
of the agent);

\item  $G_{init} \subseteq G_{vo}$;
\item  $G_{vo} = \bigcup_{\langle i, R_i, G_i \rangle \in A_{vo}} G_i$;
\item $Wf_{vo}$
is the result of instantiating $Wf$ by the given sequence of successful
dialogues, 
as dictated by $G_{vo}$;
the providers of the
services are given by $A_{vo}$.
\end{itemize}

Intuitively, agents may decide to add goals at this stage to avoid
agreements to provide a service which could prevent the fulfillment of some of their goals.
We impose that the initiating agent is not allowed to
change the workflow. However, it can add constraints or services to it,
as soon as no new role is required by this addition.
For example, this would be needed and useful to support shimming\footnote{Informally, shimming is the introduction of a service into a workflow to ensure that the output of a preceding service matches the type required by the input of the subsequent service.} of services.
Goals of provider agents may render this shimming necessary
(e.g. because a service provider does not want to interface
to another service provider).

Another aspect of this formulation is that one single provider per service is required.
These providers will need to be selected amongst all agents
that have successfully completed a dialogue with \init.
We do not impose any constraints on how this choice is performed:
the given protocols may typically dictate this.

\section{Conclusions}
\label{sec:conc}

We have described a formalisation for VOs in grid and service-oriented architectures,
formed from agent societies,
using a realistic scenario for illustration throughout.
This formalisation is abstract and independent of any realisation choices
(in terms of agent architectures, communication platform etc).
It can guide the development of (agent-based) VOs,
in that it identifies essential components (such as 
several underlying languages for services, identifiers, communication, as well as 
protocol-based roles for negotiation of services and contracts). We have experimented with some of the interactions presented here for the earth observation scenario~\cite{DBLP:conf/atal/BromuriUMST09} with emphasis on the coordination patterns agents should follow when creating a VO~\cite{mage}.

Our emphasis on the use of protocols to support VOs
is influenced by \cite{FosterKessel01}. 
The CONOISE-G~\cite{patel-2005} project
presented an agent-based model for VOs on the grid,
but focused on the challenging task of engineering a working system and thus making concrete realisation choices (e.g. agents use a constraint satisfaction algorithm for decision-making).   
We have taken a more abstract view of agents, agent society and VOs, 
to ensure that the definitions
can be  ported to any other agent-enabled grid systems to support
VOs in general.
Papers such as~\cite{Matos} speculate on the consequences of introducing software agents as a means to alleviate the burden on human decision-making.
We have a similar focus in that we see an opportunity in the use of the multiagent paradigm for automating parts of VOs.
There are a few papers that have 
formalised aspects of agent-enabled VOs, for example~\cite{DBLP:conf/atal/PittKSA05} look at voting protocols for VOs while~\cite{UdupiS06} focuses on the representation of contracts in VOs based on a specific commitment-based approach for them. 
We have taken a more exhaustive view by considering all components of agent-enabled VOs but more abstractly. 

There is also existing work applying agent societies and electronic institutions for VOs~\cite{prometheus}. There are a number of differences of this work from ours as follows. First, our emphasis is one the formalisation of a VO in terms of its components and the VO transitions, not on the details of the regulation of the participating agents and their behaviour. Secondly, we abstract away from methodological issues. Thirdly, we do not require a classification of the goals of agents and a focus on the capabilities of individual agents. The regulation of VOs and Electronic Institutions with emphasis on norms is also discussed in~\cite{Cardoso08}, where like here the focus is on agreed contracts about the provision of institutional services. However, we abstract away from the monitoring of VO activities and the evaluation of norms.

As future work, it will be important to further validate the proposed model with further examples, e.g. in e-business and pharma settings, 
as well as formally verifying that the VO formation model provided results in ``coherent'' VOs, namely VOs where all agents involved can fulfil their relevant goals 
as a result of the participation in the VO, given that the VO is executed as agreed.
 
\section*{Acknowledgements}
We would like to thank the anonymous referees for their comments on a previous version of this paper. The work reported here was partially funded by the Sixth Framework IST programme of the EC, under the
035200 ARGUGRID project.

\bibliographystyle{eptcs}
\bibliography{argugridbibs}

\end{document}